\documentclass[%
 aip,
 amsmath,amssymb,
 reprint,%
]{revtex4-1}
\usepackage{graphicx}
\usepackage{dcolumn}
\usepackage{bm}
\usepackage[utf8]{inputenc}
\usepackage[T1]{fontenc}
\usepackage{mathptmx}
 \usepackage{color}

\begin{document}

\preprint{AIP/123-QED}
%

\title{Collective dynamics of globally delay-coupled complex Ginzburg-Landau oscillators}
\author{Bhumika Thakur}
\email{bhumikathakur21@gmail.com}
\author{Abhijit Sen}%
\email{abhijit@ipr.res.in}
\affiliation{Institute for Plasma Research, HBNI, Bhat, Gandhinagar 382428, India}%


\date{\today}
             

%

%
%
\begin{abstract}
The effect of time-delayed coupling on the collective behavior of a population of globally coupled complex Ginzburg-Landau (GCCGL) oscillators is investigated. A detailed numerical study is carried out to study the impact of time delay on various collective states that include synchronous states, multicluster states, chaos, amplitude mediated chimeras and incoherent states. It is found that time delay can bring about significant changes in the dynamical properties of these states including their regions of existence and stability. In general, an increase in time delay is seen to lower the threshold value of the coupling strength for the occurrence of such states and to shift the existence domain towards more negative values of the linear dispersion parameter. Further insights into the numerical findings are provided, wherever possible, by exact equilibrium and stability analysis of these states in the presence of time delay.
\end{abstract}
%
%
\maketitle
\begin{quotation}
The Complex Ginzburg-Landau equation (CGLE) is an amplitude equation describing the dynamics of extended systems near a Hopf bifurcation and was first derived by Newell and Whitehead in 1969\cite{Newell1969}. CGLE is one of the most studied nonlinear equation in physical sciences since it and its various forms can describe a variety of phenomena such as nonlinear waves, binary fluid convection, and superconductivity. The discretized form of CGLE on a fully connected lattice is equivalent to a system of identical limit-cycle oscillators with global or all-to-all coupling \cite{Hakim1992}. Depending on the region of the parameter space, this globally-coupled system shows a variety of collective states, such as synchronization, clustering, chimeras, chaos and incoherence. Past studies on this system have been restricted to the instantaneous interaction between the oscillators. How the interaction delays affect the dynamics of such systems of globally coupled Ginzburg-Landau oscillators remains to be studied. In this paper, we address this question with a detailed numerical and analytical study.
 We find that time delay  has a significant effect on the existence and stability of the collective states of the system. Time delay is found to shift the threshold of these states towards lower values of coupling strength and more negative values of the linear dispersion parameter.
\end{quotation}
\section{Introduction}
A model system consisting of a coupled set of complex Ginzburg-Landau oscillators and its variants have been widely used as a paradigm for the mathematical study of a large variety of nonlinear phenomena in physical, chemical and biological systems \cite{Willaime1991a,Willaime1991b,Hakim1992,Nakagawa1993,Nakagawa1994,Nakagawa1995,Falcke1995,Hakim1994,Chabanol1997,
Nakao1999,Battogtokh2000,Sethia2013,Sethia2014,Zakharova2016}. One of the reasons behind the popularity and extensive applicability of this model is the great variety of collective behavior exhibited by this system ranging from synchrony to chaos and many intermediate states between them. A systematic and in-depth study of many of these collective states was made by Hakim and Rappel \cite{Hakim1992} who showed that the system can give rise to the following collective states:  homogeneous limit cycles, incoherence (a state with complete frequency locking but no phase locking), clustering and chaotic states \cite{Hakim1992}. 
Soon thereafter, Nakagawa and Kuramoto demonstrated the existence of different forms of collective chaos in this system, such as a low-dimensional collective dynamics arising from the coupled motion of three point clusters and a high-dimensional chaotic motion exhibited by fused clusters \cite{Nakagawa1993,Nakagawa1994,Nakagawa1995}.
Around the same time, Falcke et al. observed standing wave solutions with an intrinsic wavelength in such systems\cite{Falcke1995}. The effect of adding noise on the collective dynamics of this system was studied by Hakim, Rappel and Chabanol \cite{Hakim1994,Chabanol1997}. Turbulent behavior was also studied in a variant of the model where the oscillators were coupled in a non-local fashion \cite{Nakao1999,Battogtokh2000}. More recently this system was also shown to exhibit exotic states such as amplitude-mediated chimeras \cite{Sethia2013,Sethia2014} and amplitude chimeras \cite{Zakharova2016}.\\
Most of the past studies on globally coupled Ginzburg-Landau oscillators have considered the interaction between the oscillators to be instantaneous. Since time delay arising from finite propagation speed of signals is inevitable in most real-life systems, it is important to assess the influence of time delay on these states.
Our present paper is devoted to such an investigation.
Our detailed numerical investigations show that time delay has indeed a significant impact on the characteristic properties of the collective modes. In general an increase in time delay is found to lower the threshold value of the coupling strength for the occurrence of such states and to shift the existence domain towards more negative values of the linear dispersion parameter. We also determine the stability properties of these modes using a combination of numerical and analytical methods. In the limit of a small time delay (compared to the intrinsic frequency of the oscillators), the model system is shown to acquire nonlinear contributions in the coupling mechanism that enables a comparison with past model studies where such couplings have been adopted in an ad-hoc fashion.\\
The paper is organized as follows. In Sec.~\ref{sec:model}, we give the model equations and discuss the existence domain of the collective states of non-delayed system. In Sec.~\ref{sec:results}, we study the effect of delay on various collective states in detail. We show in Sec.~\ref{sec:small_delay} that the presence of a small delay has an effect similar to that of adding a nonlinear global coupling term. In Sec.~\ref{sec:discussion} we summarize our results with some concluding remarks.
\section{Model}\label{sec:model}
We study a system of $N$ globally delay-coupled complex Ginzburg-Landau oscillators governed by the following set of equations,
%
\begin{eqnarray}\label{eq:mod}
\dot{W}_j(t)=W_j(t)-(1+i C_2)|W_j(t)|^2 W_j(t)+K(1+i C_1)\;\;\nonumber\\
\left[(\overline{W}(t-\tau)-W_j(t))-\frac{1}{N}(W_j(t-\tau)-W_j(t))\right],\;\;\;\;
\end{eqnarray}
where $j=1,...,N$ and the mean field $\overline{W}=(1/N)\sum_{n=1}^{N}W_n$. 
The last term on the right hand is introduced to remove the self-coupling component that exists in the mean field summation.
Here $W_j(t)$ is the complex amplitude and $C_1$, $C_2$, and $K$ are real constants characterizing the linear and the nonlinear dispersion and the coupling strength, respectively.
The parameter $\tau$ represents the time delay in the interactions between the oscillators. 

\begin{figure}[ht!]
     \centering
\includegraphics[width=0.49\textwidth]{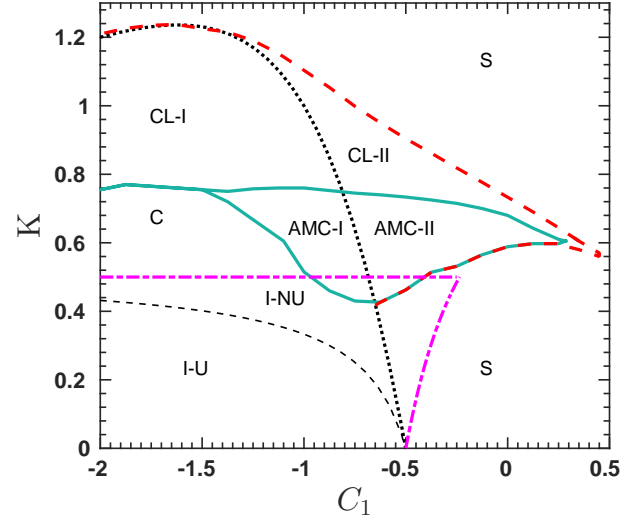} 
\caption{Phase diagram in $C_1-K$ space for $C_2=2$ in the absence of delayed interactions (i.e., $\tau=0$) showing the region of existence of various collective states. S denotes the region of stability of synchronous state. CL-I is the region of existence of multicluster states and CL-II is the region of co-existence of multiclusters and synchronous state. C denotes the region of existence of chaotic states. AMC-II denotes the region where amplitude-mediated chimera states coexist with synchronous states. AMC-I denotes the region of existence of amplitude-mediated chimeras where synchronous states are unstable. I-U denotes the stability region of incoherent states with uniform distribution of phases and I-NU denotes the incoherent states with non-uniform distribution of phases. For numerical simulations, the initial condition is a splay state and $N=201$.}
  \label{fig:phase_plt_tau0}
  \end{figure}

In the absence of time delay, previous studies have shown that depending on the value of the parameters $C_1$, $C_2$, and $K$, the system exhibits several different regimes \cite{Hakim1992,Nakagawa1993, Nakagawa1994,Chabanol1997,Sethia2014}
which include incoherent state, chaotic state, amplitude mediated chimera (AMC), multicluster states, and synchronous state. 
The regions of their existence and stability in $K-C_1$ phase space for a fixed value of $C_2=2$ are shown in Fig.~\ref{fig:phase_plt_tau0}. We now briefly describe this figure and discuss the details of some of these past results obtained in the absence of delay.\\

The stability criteria for the synchronous state and the incoherent state can be obtained analytically\cite{Hakim1992,Nakagawa1993,Nakagawa1994,Chabanol1997}.
In synchronous state all the oscillators are in the same state and follow a limit cycle given by $W_j=\exp(-i C_2 t)$ $\forall j$. 
Therefore, in the complex $W$ plane, oscillators move along a circle of radius unity with angular velocity $-C_2$.
A linear stability analysis of the synchronous solution gives the stability condition\cite{Hakim1992,Nakagawa1994}
\begin{equation}\label{eq:tau0_sync}
(1+C_1^2)K+2(1+C_1 C_2)>0.
\end{equation}
The marginal stability curve for the synchronous state obtained from Eq.~\eqref{eq:tau0_sync}  is shown by the black dotted line in Fig.~\ref{fig:phase_plt_tau0}. The synchronous state is stable in the region denoted by $S$ to the right of the curve.

In the incoherent state the phases of the oscillators are distributed on a circle of radius $\sqrt{1-K}$ in such a way that the mean field vanishes. The incoherent solution is given by $W_j=\sqrt{1-K} \exp\left(-i\left[K C_1 +(1-K) C_2\right]t+i \phi_j\right)$, such that $\sum_{j=1}^N \exp(i\phi_j)/N=0$. A linear stability analysis of this incoherent solution gives the stability condition\cite{Nakagawa1994,Chabanol1997}
\begin{eqnarray}\label{tau0_inc}
\left[(2-3 K)^2 + C_1^2 K^2\right] \left[4(K-1)(2K-1) C_1C_2\right.\nonumber\\
\left.+ K (2 K -1)C_1^2-K (K-1) C_2^2+(2-3K)^2 \right] \nonumber\\
+\Delta^2 K(K-1)(2-3 K)^2(1+C_1^2)(1+C_2^2) <0.
\end{eqnarray}
The marginal stability curve of incoherent states with uniform distribution of phases (also known as splay states) is obtained by taking $\Delta=0$ in Eq.~\eqref{tau0_inc}. This curve is shown by a black dashed curve in Fig.~\ref{fig:phase_plt_tau0}. Incoherent states having non-uniform distribution of phases have a larger region of stability with the most stable state corresponding to $\Delta=1$ in Eq.~\eqref{tau0_inc} \cite{Hakim1992}. The stability boundary of these states is denoted by the magenta dashed dotted curve in Fig.~\ref{fig:phase_plt_tau0}. 

The stability regions of the multi-cluster states, the chaotic state and the AMC have been obtained numerically for $N=201$ oscillators for an initial condition that is a splay state. 
In a multi-cluster state, the oscillator population splits into two or three clusters such that the oscillators within a cluster are in the same state but the oscillators in different clusters are in different states. 
The stability region of these states is denoted by CL-II and CL-I in Fig.~\ref{fig:phase_plt_tau0} depending on whether they co-exist with the synchronous state or not.
The dynamical behavior in a two-cluster state can be periodic or quasi-periodic depending on parameter values \cite{Nakagawa1993}. The dynamics in a three-cluster state is low-dimensional chaotic. High-dimensional chaotic states arise from the merger of the three point-clusters into a continuous distribution in the complex W-plane.
Chaotic states are stable in the region marked C between the incoherent states and the multicluster states. In a chaotic state the behavior of each oscillator is quite complex but the oscillators maintain some level of coherence such that mean field does not vanish.

For some parameter values, only two clusters fuse together to form a continuous distribution while the third point-cluster maintains its form. This coexistence of continuous distribution and point cluster has been termed as an amplitude-mediated chimera state.
The region bounded by the greenish-blue solid curve in Fig.~\ref{fig:phase_plt_tau0} is the region of existence of AMC states. 
Chimera states represent a spontaneous splitting of a population of oscillators into sub-populations displaying synchronized and de-synchronized behavior.  
Chimera states were first discovered in a system of coupled phase oscillators by Kuramoto and Battogtokh \cite{Kuramoto2002}. For nearly a decade the research on chimera states focused solely on phase-only chimera states.
AMCs were first noticed by Nakagawa and Kuramoto\cite{Nakagawa1993} in globally coupled complex Ginzburg-Landau oscillators but they remained unidentified and ignored for two decades. 
They were studied for the first time by Sethia and Sen initially for nonlocally\cite{Sethia2013} and later for globally \cite{Sethia2014} coupled Ginzburg-Landau oscillators.
AMCs are named so because they show chimeric behavior in both amplitude as well as phase variables.
In the globally coupled system, they are found to be stable in some region of the parameter space between the three-cluster state and chaos which is denoted by AMC-I in Fig.~\ref{fig:phase_plt_tau0}. They are also found to co-exist with synchronous states in the region marked AMC-II in Fig.~\ref{fig:phase_plt_tau0}.

In the following section, we study the effect of time-delayed interactions on the existence and stability of all these collective states.
\section{Results in the presence of time delay}\label{sec:results}
Our investigations on the effect of time delay on the collective dynamics of Eq.~\eqref{eq:mod} have mostly been carried out with the help of numerical simulations of the system (\ref{eq:mod}). Wherever possible, numerical results are supported with exact equilibrium and stability analysis.
The simulations have been carried out mostly with $N=201$ globally delay-coupled discrete oscillators using MATLAB's delay differential equation solver dde23. 
The various collective states are discussed in the sequence of their emergence with decreasing coupled strength $K$.
\subsection{Synchronous (single-cluster) state}
When the coupling strength $K$ is sufficiently high, all the oscillators are perfectly synchronized and therefore, form a single point-cluster.
Since all oscillators are in the same state, i.e., $W_j (t)=W(t)$, Eq.~\eqref{eq:mod} becomes
\begin{eqnarray}\label{eq:1c1}
\dot{W}(t)=W(t)-(1+i C_2)|W(t)|^2 W(t)+K(1-1/N)\nonumber\\(1+i C_1)(W(t-\tau)-W(t)).\;\;\;
\end{eqnarray}
Let us take $d=1-1/N$ to re-write the above equation as
\begin{eqnarray}\label{eq:1c2}
\dot{W}(t)=W(t)-(1+i C_2)|W(t)|^2 W(t)+K d (1+i C_1)\nonumber\\(W(t-\tau)-W(t)).\;\;\;
\end{eqnarray}
For large number of oscillators, $d\approx 1$. The stable synchronous solution corresponds to oscillators following a limit cycle with $W(t) = A e^{i\Omega t}$, where $A$ is the amplitude and $\Omega$ is the collective frequency. Substituting this solution in Eq.~\eqref{eq:1c2} and separating the real and imaginary parts gives the set of coupled equations 
\begin{equation}\label{eq:1c3}
 A^2=1+Kd\left(\cos[\Omega\tau]-1 + C_1 \sin[\Omega\tau]\right),
\end{equation}
and
\begin{equation}\label{eq:1c4}
 \Omega=-C_2 A^2 -K d\left( \sin[\Omega\tau] -C_1 \left(\cos[\Omega\tau]-1\right)\right).
\end{equation}
Substituting the expression for $A^2$ from Eq.~\eqref{eq:1c3} into Eq.~\eqref{eq:1c4} gives the following transcendental equations for $\Omega$ 
\begin{eqnarray}\label{eq:1c5}
 \Omega = - C_2 + K d\left(  (C_2-C_1) \left(1-\cos[\Omega\tau]\right) \right.\nonumber\\
\left. -(1+C_1 C_2)\sin[\Omega\tau]\right).
\end{eqnarray}
The expressions (\ref{eq:1c3}) and (\ref{eq:1c5}), for $A$ and $\Omega$ respectively, show that the amplitude and frequency depend on the delay parameter $\tau$. In the absence of delay ($\tau=0$), $A=1$, i.e., the oscillators run around a unit circle at a constant frequency $\Omega=-C_2$.
\medskip\\
Next we find the region of stability of the single-cluster synchronous state. 
Eq.~\eqref{eq:mod} is linearized by substituting $W_j(t)=(A+a_j(t)) e^{i\Omega t}$, where $a_j$ is a complex perturbation to the state $W_j$ such that $|a_j|<<1$.
The linear evolution of the perturbation $a_j$ is given by
\begin{eqnarray}\label{eq:stab_1C_1}
 \frac{d a_j(t)}{dt}=-K (1+i C_1) e^{-i \Omega \tau} \left(d +\frac{e^{-\lambda \tau}}{N}\right) a_j(t)\nonumber\\
 -(1+i C_2) A^2 (a_j(t)+a_j^* (t))\nonumber\\
 +K (1+i C_1) e^{-i \Omega \tau}e^{-\lambda \tau}\frac{1}{N}\sum_k a_k(t),
 \end{eqnarray}
where $^*$ denotes complex conjugation. The corresponding $2N\times2N$ stability matrix $M$ has a large symmetry and can be written as 
\[
 M=\begin{bmatrix}
    m_1 & m_2 & m_3 & 0 & m3 & 0&\cdots & m_3 & 0\\
    m_4 & m_5 & 0 & m_6 & 0 & m_6 &\cdots & 0 & m_6\\
    m_3 & 0 & m_1 & m_2 & m_3 & 0&\cdots & m_3 & 0\\
    0 & m_6 & m_4 & m_5 & 0 & m_6&\cdots& 0 & m_6\\
    m_3 & 0 & m_3 & 0 & m_1 & m_2&\cdots & m_3 & 0 \\
    0 & m_6 & 0 & m_6 & m_4 & m_5 &\cdots& 0 & m_6\\
    \vdots & \vdots & \vdots & \vdots & \vdots & \vdots &\ddots & \vdots & \vdots \\
     m_3 & 0 & m_3 & 0 & m_3 & 0&\cdots & m_1 & m_2\\
    0 & m_6 & 0 & m_6 & 0 & m_6 &\cdots& m_4 & m_5\\
  \end{bmatrix},
\]
where,
\begin{eqnarray}
 m_1 &=& - K d (1 + i C_1)e^{-i\Omega\tau}- A^2 (1 + i C_2), \nonumber\\
 m_2&=&- A^2 (1 + i C_2), \nonumber\\
 m_3&=&(K/N) (1 + i C_1)e^{-\lambda\tau-i\Omega\tau}, \nonumber\\
 m_4&=&m_2^*=- A^2 (1 - i C_2), \nonumber\\
 m_5&=&m_1^*=- K d (1 - i C_1)e^{i\Omega\tau}- A^2 (1 - i C_2), \; \text{and} \nonumber\\
 m_6&=&(K/N) (1 - i C_1)e^{-\lambda\tau+i\Omega\tau}. \nonumber
\end{eqnarray}

The characteristic eigenvalue equation is given by
$|M-\lambda I |=0$, where $I$ is the $2N\times2N$ identity matrix. The eigenvalue equation can be written as
\begin{align}
(-1)^N \left[\underbrace{m_2 m_4+(m_1- m_3-\lambda)(-m_5+m_6+\lambda)}_{M_1(\lambda)}\right]^{N-1}\;\;\;\;\;\;\;\;\;\; \nonumber\\
\underbrace{(m_2 m_4-(m_1+(N-1)m_3-\lambda)(m_5+(N-1)m_6-\lambda))}_{M_2(\lambda)}\;\;\;\;\;\;\;\;\;\;\;\nonumber\\
=0.\;\;\;\;\;\;\;\;\;\;\;\;\;\;\;\;
\end{align}
This gives $N-1$ times the couple of eigenvalues $(\lambda_1,\lambda_2)$ obtained from $M_1(\lambda)=0$. 
The other two eigenvalues, one of which is always $0$, are obtained from taking $M_2(\lambda)=0$. 
The zero eigenvalue comes from the invariance of Eq.~\eqref{eq:mod} under a global phase change \cite{Chabanol1997}.

\begin{figure}[ht!]
     \centering
\includegraphics[width=0.48\textwidth]{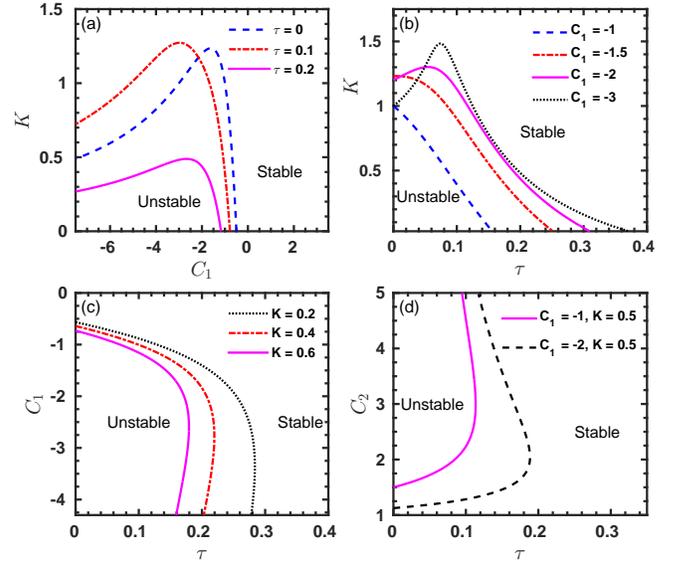}
\caption{The marginal stability curves for the one-cluster synchronous state in (a) $K - C_1$ parameter space at $C_2=2$ for various values of $\tau$, (b) $K-\tau$ parameter space at $C_2=2$ for various values of $C_1$, (c) $C_1-\tau$ parameter space at $C_2=2$ for various values of $K$, and (d) $C_2-\tau$ parameter space at $K=0.5$ for two values of $C_1$.}
   \label{fig:1c_stab} 
  \end{figure}

On taking $\tau=0$, we get the eigenvalue equation for the non-delayed case \cite{Hakim1992,Chabanol1997} as
\begin{eqnarray}\label{eq:stab_1c_4a}
\lambda^2+2(K+1)\lambda+ (1+C_1^2)K^2+2 K\left(1+C_1 C_2\right)=0,\;\;\;\;\;\;\; 
\end{eqnarray}
and the marginal stability curve is given by Eq.~\eqref{eq:tau0_sync}.
In Fig.~\ref{fig:1c_stab}, we have plotted the stability curves for the homogeneous one-cluster state in various parameter spaces. 
Fig.~\ref{fig:1c_stab}(a) shows the stability curves in $K-C_1$ parameter space at a fixed value of $C_2=2$ for various values of $\tau$. The one-cluster state is stable in the region above the curves.
The region of stability shifts towards more negative values of $C_1$ with an increase in delay.
Fig.~\ref{fig:1c_stab}(b) shows stability curves in $K-\tau$ parameter space for various values of $C_1$. For a given value of $C_1$, the critical value of $K$ at which the one-cluster state becomes stable decreases with increasing delay.  
Fig.~\ref{fig:1c_stab}(c) shows stability curves in $C_1-\tau$ parameter space for various values of $K$ and we find that with increasing delay the stability region extends to more negative values of $C_1$. Also, the stability region expands with an increase in $K$. 
Fig.~\ref{fig:1c_stab}(d) shows the stability curves in $C_2-\tau$ parameter space. 
For a fixed value of $K$ and $C_1$, the unstable region lies in the parameter space with large values of $C_2$ and small values of delay.
Overall, the delay parameter seems to favor the stabilization of the single-cluster state.   
\subsection{Clustering}
As one decreases the coupling strength $K$ below a critical value, the system shows clustering behavior, where the single synchronous cluster breaks down into two or three groups or clusters of identical oscillators.
Clustering behavior in GCCGLE in the absence of delay has been reported previously by Nakagawa and Kuramoto \cite{Nakagawa1993} and recently in ensembles of Stuart-Landau oscillators \cite{Kemeth2018}.
\subsubsection{Two-cluster state}
When $K$ decreases below a certain value, the system first makes a transition from the single-cluster state to the two-cluster state where the oscillators within each cluster have same amplitude and phase but these are different for the two clusters. For a range of $K$, the single-cluster and two-cluster states also co-exist.
In a two-cluster state, the dynamics of the oscillators can be periodic or quasi-periodic.
In the periodic (homogeneous limit cycle) state, all the oscillators are frequency synchronized. The oscillators within a cluster are synchronized in-phase, while there is a finite phase difference between the two clusters. The coupled equations for the amplitudes and frequency can be determined by substituting the limit cycle solutions for the two clusters.
Suppose $pN$ number of oscillators with complex amplitude $W_1(t)$ are in one cluster, and the remaining $(1-p) N =qN$ number of oscillators with complex amplitude $W_2(t)$ are in second cluster.
Then the $N$ coupled equations of motion reduce to two coupled equations that give the evolution of two point-clusters.
Thus we have,
\begin{eqnarray}\label{eq:2c1}
\dot{W_1}(t)=W_1(t)-(1+i C_2)|W_1(t)|^2 W_1(t)+K(1+i C_1) \nonumber\\
\left[p'\; W_1(t-\tau)+q\; W_2(t-\tau) - d\; W_1(t)\right],\;\;\;\;\;\;
\end{eqnarray}
\begin{eqnarray}\label{eq:2c2}
\dot{W_2}(t)=W_2(t)-(1+i C_2)|W_2(t)|^2 W_2(t)+K(1+i C_1)\nonumber\\
\left[q'\; W_2(t-\tau) + p\; W_1(t-\tau) - d\; W_2(t)\right], \;\;\;\;\;\;
\end{eqnarray}
where $p'=p-1/N$ and $q'=q-1/N$.
Let us assume separate limit-cycle solutions for the two point-clusters. We take $W_1(t)=A_1 e^{i\left(\Omega t+\phi_1\right)}$, where $\Omega$ is the collective frequency and $\phi_1$ is the phase offset of the oscillators in the first cluster, and $W_2(t)=A_2 e^{i\left(\Omega t+\phi_2\right)}$, where $\phi_2$ is the phase offset of the oscillators in the second cluster.
Substituting these in Eqs.~(\ref{eq:2c1},\ref{eq:2c2}) and separating the real and imaginary parts gives four coupled equations
\begin{eqnarray}\label{eq:2c3}
A_1^2&=&1-K d+K p'\left(\cos[\Omega\tau]+C_1\sin[\Omega\tau]\right)\nonumber\\
&+&K q \frac{A_2}{A_1}\left(\cos[\Delta \phi-\Omega\tau]-C_1\sin[\Delta \phi-\Omega\tau]\right),
\end{eqnarray}
\begin{eqnarray}\label{eq:2c4}
\Omega&=& -C_2 A_1^2-C_1 K d-K p'\left(\sin[\Omega\tau]-C_1\cos[\Omega\tau]\right)\nonumber\\
&+& K q \frac{A_2}{A_1}\left(\sin[\Delta \phi-\Omega\tau]+C_1\cos[\Delta \phi-\Omega\tau]\right),
 \end{eqnarray}
\begin{eqnarray}\label{eq:2c5}
A_2^2&=&1-K d+K q'\left(\cos[\Omega\tau]+C_1\sin[\Omega\tau]\right)\nonumber\\
&+&K p \frac{A_1}{A_2}\left(\cos[\Delta \phi+\Omega\tau]+C_1\sin[\Delta \phi+\Omega\tau]\right),
 \end{eqnarray}
\begin{eqnarray}\label{eq:2c6}
\Omega &=& -C_2 A_2^2-C_1 K d-K q'\left(\sin[\Omega\tau]-C_1\cos[\Omega\tau]\right)\nonumber\\
&-& K p \frac{A_1}{A_2}\left(\sin[\Delta \phi+\Omega\tau]-C_1\cos[\Delta \phi+\Omega\tau]\right),
\end{eqnarray}
where $\Delta \phi=\phi_2-\phi_1$.
The collective frequency $\Omega$ is non-isochronous, i.e.,  it depends on the amplitudes of the oscillators in each cluster. It also depends on the fractions, $p$ and $q=1-p$, of oscillators in each cluster.
Here, we have four coupled equations and five quantities to be determined - $A_1$, $A_2$, $\Omega$, $\Delta \phi$, and $p$. For a given value of one of the variables, these coupled equations give the values of the remaining four which are found to be in agreement with their values obtained from numerical integration of Eq.~\eqref{eq:mod}.

\begin{figure}[ht!]
     \centering
\includegraphics[width=0.49\textwidth]{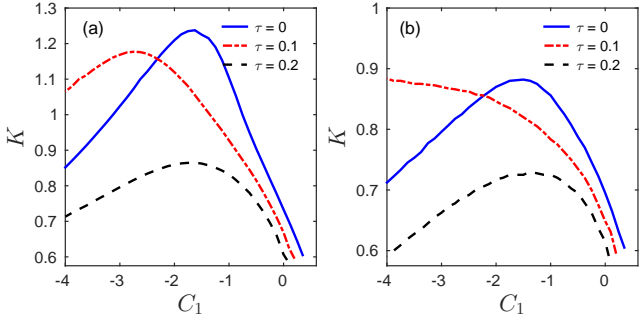}
     \caption{Numerically obtained stability curves for (a) upper bound, and (b) lower bound of the two-cluster periodic state for an initial condition that is a splay state. The curves are plotted for three values of delay parameter $\tau=0$ (solid blue line), $\tau=0.1$ (red dashed-dotted line), and $\tau=0.2$ (black dashed line). Other parameters are: $C_2=2$ and $N=201$.}
  \label{fig:plt_2c}
  \end{figure}

To determine whether the values of amplitudes and frequency obtained from these coupled equations correspond to a stable two-cluster periodic state, we perform a linear stability analysis. We substitute in Eq.~\eqref{eq:mod} the perturbed form of the solutions, $W_{j_1}(t)=(A_1+a_{j_1}(t)) e^{i \left(\Omega t+ \phi_1\right)}$ and
$W_{j_2}(t)=(A_2+a_{j_2}(t)) e^{i \left(\Omega t+ \phi_2\right)}$, where $j_1=1,...,p N$ and $j_2=pN+1,...,N$. We get the following equations for the evolution of the perturbations $a_{j_1}(t)$ and $a_{j_2}(t)$.
\begin{eqnarray}
 \frac{d a_{j_1}}{d t}=K (1+i C_1) e^{-i \Omega \tau} \left[-p'-\frac{e^{-\lambda\tau}}{N}-q \frac{A_2}{A_1} e^{i \Delta\phi}\right] a_{j_1}(t)\nonumber\\
-(1+i C_2) A_1^2\left(a_{j_1}(t)+a_{j_1}^*(t)\right)+K (1+i C_1)e^{-i \Omega \tau } \nonumber\\
 \times e^{-\lambda \tau} \frac{1}{N}\left[\sum_{k_1=1}^{p N} a_{k_1}(t)+\sum_{k_2=p N +1}^{ N} a_{k_2}(t)e^{i \Delta \phi}\right].\;\;\;\;\;\;\;
 \end{eqnarray}
\begin{eqnarray}
 \frac{d a_{j_2}}{d t}=K (1+i C_1) e^{-i \Omega \tau} \left[-q'-\frac{e^{-\lambda\tau}}{N}-p \frac{A_1}{A_2} e^{-i \Delta\phi}\right] a_{j_2}(t)\nonumber\\
-(1+i C_2) A_2^2\left(a_{j_2}(t)+a_{j_2}^*(t)\right)+K (1+i C_1)e^{-i \Omega \tau } \nonumber\\
 \times e^{-\lambda \tau} \frac{1}{N}\left[\sum_{k_1=1}^{p N} a_{k_1}(t) e^{- i \Delta \phi} +\sum_{k_2=p N +1}^{ N} a_{k_2}(t)\right].\;\;\;\;\;\;\;
 \end{eqnarray}
The stability matrix of dimensions $2N\times2N$ of the perturbations $a_{\{j_1,j_2\}}$ and their complex conjugates $a_{\{j_1,j_2\}}^*$ has a lot of symmetry. Therefore, it is possible to obtain the characteristic eigenvalue equation 
 which can be written as
\begin{eqnarray}\label{eq:2c_eg}
\left[ \left(N_1 N_2 l_1 l_2^* s_2 s_3 +|l_1|^2 \left(-|l_2|^2 + \left(h_2+(N_2-1) n_1-\lambda\right)\right.\right.\right.\;\;\;\;\;\; \nonumber\\
\left.\left.\left(h_2^*+(N_2-1)n_2-\lambda\right)\right)\right)+\left(N_1 N_2 l_2 l_1^* s_1 s_4+\;\;\;\;\;\;\right.\nonumber\\
\left.|l_2|^2\left(h_1+(N_1-1)n_1-\lambda\right)\left(h_1^*+(N_1-1)n_2-\lambda\right)\right)-\;\;\;\;\;\;\nonumber\\
\left(N_1 N_2 n_1^2 -\left(h_1+(N_1-1)n_1-\lambda\right)\left(h_2+(N_2-1)n_1-\lambda\right)\right)\;\nonumber\\
\left.\left(N_1 N_2 n_2^2 -\left(h_1^*+(N_1-1)n_2-\lambda\right)\left(h_2^*+(N_2-1)n_2-\lambda\right)\right)\right]\;\nonumber\\
\left(|l1|^2+(h_1-n_1-\lambda)(-h_1^*+n_2+\lambda)\right)^{N_1-1}\;\;\;\;\;\;\nonumber\\
\left(-|l_2|^2-(h_2-n_1-\lambda)(-h_2^*+n_2+\lambda)\right)^{N_2-1}=0,\;\;\;\;\;\;
\end{eqnarray}
where $^*$ denotes complex conjugation. $N_1=p N$ is the number of oscillators in first cluster and $N_2=q N$ is the number of oscillators in the second cluster;
\begin{eqnarray}
h_1&=&  K (1 + i C_1)e^{-i\Omega\tau}\left(-p'-q \frac{A_2}{A_1} e^{i \Delta\phi}\right)- A_1^2 (1 + i C_2), \nonumber\\
h_2&=&  K (1 + i C_1)e^{-i\Omega\tau}\left(-q'-p \frac{A_1}{A_2} e^{-i \Delta\phi}\right)- A_2^2 (1 + i C_2), \nonumber\\
n_1&=& K (1 + i C_1)e^{-i\Omega\tau}e^{-\lambda\tau}/N, \nonumber\\
n_2&=& K (1 - i C_1)e^{i\Omega\tau}e^{-\lambda\tau}/N, \nonumber\\
l_1&=&- A_1^2 (1 + i C_2),\nonumber\\
l_2&=&- A_2^2 (1 + i C_2),\nonumber\\
s_1&=& n_1  e^{i \Delta\phi},\nonumber\\
s_2&=& n_2  e^{-i \Delta\phi},\nonumber\\
s_3&=& n_1  e^{-i \Delta\phi},\; \text{and}\nonumber\\
s_4&=& n_2  e^{i \Delta\phi}. \nonumber
\end{eqnarray}
All the eigenvalues obtained from Eq.~\eqref{eq:2c_eg} for a given set of parameters should have a non-positive real part for the corresponding two-cluster periodic state to be stable.
In Fig.~\ref{fig:plt_2c}, we have plotted the numerically obtained upper-bound (Fig.~\ref{fig:plt_2c}(a)) and lower-bound (Fig.~\ref{fig:plt_2c}(b)) stability curves in the $K-C_1$ parameter space of the two-cluster periodic state. The curves are plotted for three values of delay parameter $\tau=0$ (solid blue curve), $0.1$ (red dashed-dotted curve), and  $0.2$ (black dashed curve)  for an initial condition that is a splay state. 
The numerical results indicate that the stability domain of two-cluster periodic state shifts towards lower values of coupling parameter with an increase in delay.
The stability domain of the two-cluster periodic state and the number of oscillators in either cluster also depend on the choice of initial condition and the value of other parameters.

For a given delay, as we decrease the coupling strength below a certain value the dynamic behavior of the two-cluster state changes from periodic to quasi-periodic.
\subsubsection{Three-cluster state}
On further decreasing the coupling strength, the smaller cluster of the two-cluster state further splits into two clusters and we get three point-clusters. Unlike the two-cluster state, the three-cluster state does not seem to show periodic behavior where oscillators in each cluster have a homogeneous limit cycle. Instead, the collective dynamics of the three point-clusters is (low-dimensional) chaotic as has previously been observed in the non-delayed case \cite{Nakagawa1993}.
\subsubsection{Location and sizes of clusters}
\begin{figure*} 
     \centering
\includegraphics[width=\textwidth,height=4in]{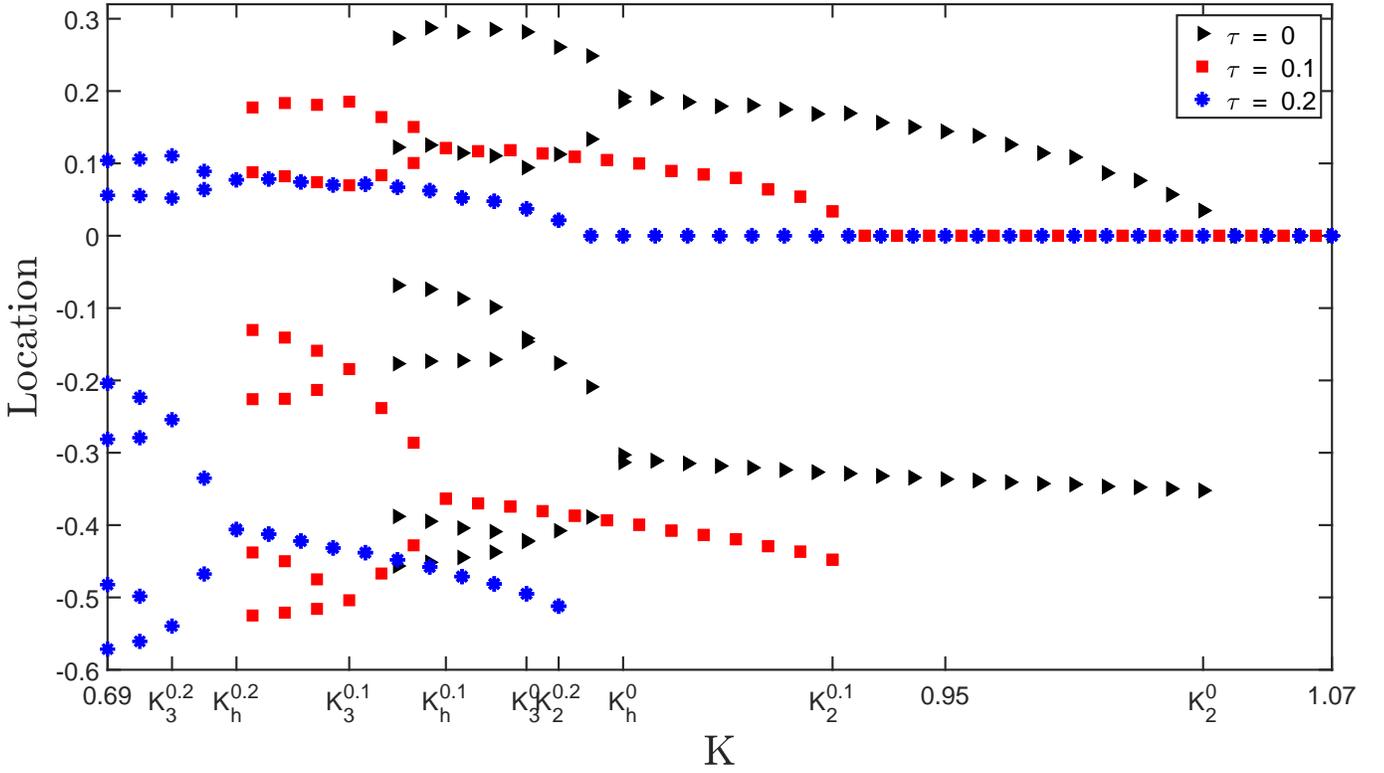}
     \caption{Locations of the clusters (Im($Z_j$)) are plotted as a function of the coupling parameter $K$ for three values of time delay parameter $\tau=0$ (black triangles), $\tau=0.1$ (red squares), and $\tau=0.2$ (blue stars). Other parameters are fixed at $C_1=-1$, $C_2=2$ and $N=100$. In each case, above $K_2^{\tau}$, one-cluster state (i.e., the synchronous state) is stable. As $K$ is decreased below $K_2^{\tau}$ the stationary two-cluster state becomes stable and remains stable till $K_h^{\tau}$ where the clusters start to oscillate. On further decreasing K, the three-cluster state becomes stable at $K_3^{\tau}$. The initial condition is a splay state.}
 \label{fig:Location_vs_k}
  \end{figure*}

\begin{figure*}
     \centering
\includegraphics[width=\textwidth,height=4in]{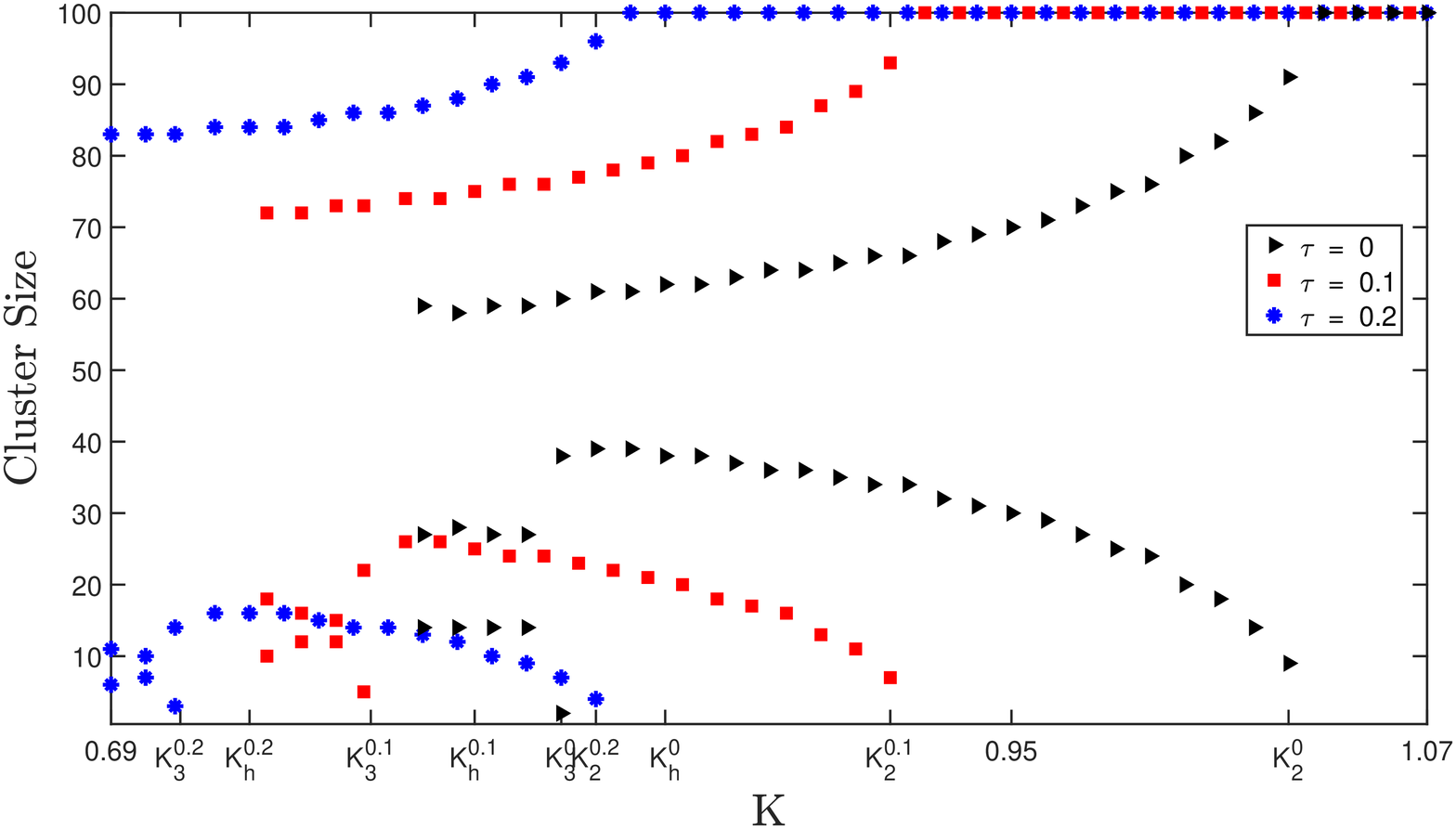}
     \caption{Cluster sizes are plotted as a function of the coupling parameter $K$ for three values of time delay parameter, $\tau=0$ (black triangles), $\tau=0.1$ (red squares), and $\tau=0.2$ (blue stars). All the parameters are same as Fig.~\ref{fig:Location_vs_k}. Above $K_2^{\tau}$, all the oscillators are part of a single large cluster of size $N=100$. Below $K_2^{\tau}$, this cluster splits into two clusters and the size of smaller cluster increases with decreasing coupling strength. On decreasing the coupling strength below $K_3^{\tau}$, the smaller cluster further splits into two clusters.}
 \label{fig:p_vs_k}
  \end{figure*}

The clustering mechanism with decreasing $K$ is part of the sequence of bifurcations that lead to collective chaos.
For understanding this mechanism for the delayed case,  we follow the analysis for the non-delayed case by Nakagawa and Kuramoto \cite{Nakagawa1993} and go to a frame in which the phase of complex mean field $\overline{W}(t)$ looks constant. We take
\begin{eqnarray}
 \overline{W}=R \exp(i\theta),\\
 W_j=Z_j \exp(i\theta),
\end{eqnarray}
where $R=\sum_j Z_j/N$. In the new frame, one-cluster state is represented by a fixed point. Using the same parameters as Nakagawa and Kuramoto \cite{Nakagawa1993}, we fix $C_1$ at $-1$ and $C_2$ at $2$. 
We plot the variation of the locations of the clusters ($\text{Im} Z_j$) and their sizes with $K$ for three values of $\tau$ in Figs.~\ref{fig:Location_vs_k} and \ref{fig:p_vs_k} respectively.
In all simulations, the initial condition is a splay state and $N=100$. As reported in the non-delayed case\cite{Nakagawa1993}, when we decrease $K$ from above, at a critical value of coupling there is a transition from one- to two-cluster state and this transition is discontinuous.
The coupling strength at which this transition takes place is denoted as $K_2^{\tau}$, where the superscript represents the corresponding value of delay. 
Above $K_2^{\tau}$, all the oscillators form a single large cluster of size $N=100$ and below $K_2^{\tau}$, this cluster splits into two clusters (see Fig.~\ref{fig:p_vs_k}). The size of smaller cluster increases with decreasing coupling strength.
From Fig.~\ref{fig:Location_vs_k}, we observe that the value of $K_2^{\tau}$ decreases as $\tau$ increases, implying that an increase in time delay decreases the threshold of stability of one-cluster state. 
As we further decrease $K$ to $K_h^{\tau}$, the stationary two-cluster state becomes oscillatory unstable, which is indicated by a pitchfork diagram in each cluster for each delay value. The two branches show the maxima and minima of the amplitude of the cluster oscillations. We observe that as we increase delay, the difference between the location of maxima and minima of the amplitude of the cluster oscillations decreases, which indicates that delay is working towards suppressing the oscillatory instability. On further decreasing $K$ down to $K_3^{\tau}$, the smaller cluster splits into two and we get three point-clusters whose coupled dynamics shows low-dimensional chaos. 
As in the case of the two-cluster state, our numerical results for the three-cluster state
also display some dependence on the initial conditions. For our comparative studies we have kept the initial
conditions to be the same for all our simulations e.g. a splay state.
The cluster sizes also depend on other parameters such as the coupling strength $K$ and time delay $\tau$.

As $K$ is further decreased the non-delayed GCCGLE can demonstrate different forms of collective chaos \cite{Nakagawa1993,Nakagawa1994}.
In some cases, the oscillators form a $\rho$-shaped curve in the phase space which rotates and deforms in time \cite{Chabanol1997,Nakagawa1994}.
Like all the other collective states, presence of delay may also influence the stability of the chaotic states - a challenging topic that has not been explored in our present study.  
\subsection{Incoherent state}
\begin{figure}[ht!]
     \centering
\includegraphics[width=0.49\textwidth]{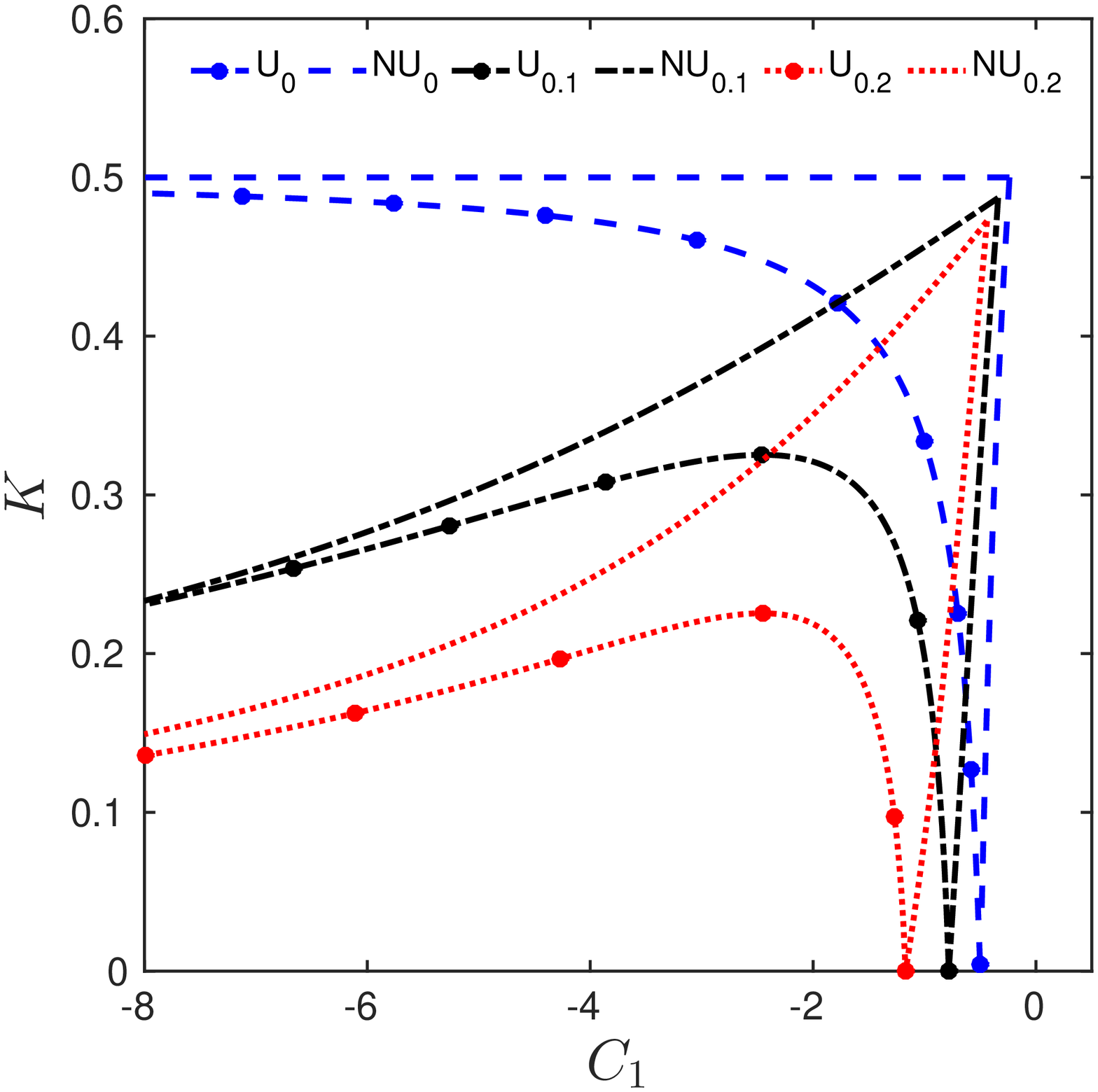} 
     \caption{Phase diagram showing the marginal stability curves for splay(uniform) and nonuniform incoherent states in $C_1-K$ space for $\tau = 0$ (blue dashed-dotted curve for uniform and dashed curve for nonuniform), $\tau = 0.1$ (black dashed-dotted curve for uniform and dashed curve for nonuniform), and $\tau = 0.2$ (red dashed-dotted curve for uniform and dashed curve for nonuniform). In the legend, $U$ denotes incoherent states that have uniformly distributed phases and $NU$ corresponds to states with nonuniformly distributed phases on a circle. Subscripts denote the corresponding value of delay. We have taken $C_2=2$ and $N=201$. For a given value of delay, the incoherent state is stable in the region below the curve and loses stability on crossing the curve into the region of higher values of $K$ and more positive values of $C_1$. The stability region decreases with increase in delay.}
 \label{fig:inc}
  \end{figure}

For small values of coupling strength $K$, the oscillators behave incoherently.
The oscillators rotate with the same frequency but are distributed on a circle in such a way that the mean field vanishes.
We will denote the frequency of the oscillators in the incoherent states as $\Omega_{inc}$ and amplitude as $A_{inc}$.
Therefore, the incoherent solution of Eq.~\eqref{eq:mod} is given by $W_j=A_{inc} e^{i(\Omega_{inc}t+\phi_j)}$, where
\begin{equation}
 A_{inc}^2=1-Kd-\frac{K}{N}\cos(\Omega_{inc} \tau)-\frac{K C_1}{N}\sin(\Omega_{inc} \tau),
\end{equation}%
\begin{eqnarray}
 \Omega_{inc}=-C_2+K d (C_2-C_1)+\frac{K (C_2-C_1)}{N}\cos(\Omega_{inc} \tau)\nonumber\\
 +\frac{K (1+C_2 C_1)}{N}\sin(\Omega_{inc} \tau).\;\;\;\;\;\;\;
\end{eqnarray}
For $\tau=0$, these expressions reduce to those derived for the non-delayed case \cite{Nakagawa1993,Chabanol1997}.
To find the criteria for the stability of the incoherent state, we perturb this solution such that
\begin{equation}\label{inc_stab1}
 W_j=(A_{inc}+a_j(t)) e^{i (\Omega_{inc}t + \phi_j)},
\end{equation}
$|a_j|\ll1$. 
On substituting this perturbed solution in Eq.~\eqref{eq:mod} and upon linearization, the equation for the perturbation amplitude $a_j$ can be written as
\begin{eqnarray}
\frac{d a_j(t)}{dt}= K\left(1+i C_1\right) e^{-i \Omega\tau} \frac{1}{N}\sum_k a_k\left(t-\tau\right)e^{i(\phi_k-\phi_j)}\nonumber\\
+\frac{K \left(1+i C_1\right)}{N} e^{-i \Omega\tau} \left[a_j(t)-a_j(t-\tau)\right]\nonumber\\
-(1+i C_2) A_{inc}^2 \left[a_j(t)+a_j^*(t)\right].\;\;\;\;
\end{eqnarray}
%
Assuming the perturbations to vary as $e^{\lambda t}$, the marginal stability equation can be written as
\begin{align}
\left[u_2 u_4+\left(u_1-u_3-\lambda\right)\left(-u_5+u_6+\lambda\right)\right]^{N-2}\nonumber\\
\bigg(\left[u_2 u_4+ \left(u_1-u_3-\lambda\right)\left(-u_5-(N-1)u_6+\lambda\right)\right] \nonumber\\
\left[u_2 u_4+ \left(u_1+(N-1)u_3-\lambda\right)\left(-u_5+u_6+\lambda\right)\right] -D \bigg)=0,
\end{align}
where
\begin{eqnarray}
 u_1 &=& (K/N) (1 + i C_1) e^{-i\Omega\tau}- A^2 (1 + i C_2), \nonumber\\
 u_2&=&- A^2 (1 + i C_2), \nonumber\\
 u_3&=&(K/N) (1 + i C_1)e^{-\lambda\tau-i\Omega\tau}, \nonumber\\
 u_4&=&- A^2 (1 - i C_2), \nonumber\\
 u_5&=&(K/N) (1 - i C_1) e^{i\Omega\tau}- A^2 (1 - i C_2),  \nonumber\\
 u_6&=&(K/N) (1 - i C_1) e^{-\lambda\tau+i\Omega\tau}, \text{and} \nonumber\\
 D&=&A_{inc}^4 (1+C_1^2)(1+C_2^2)e^{-2\lambda\tau} K^2 \Delta^2. \nonumber
\end{eqnarray}
For the splay state, i.e., when the phases are uniformly distributed on a circle, $\Delta = 0$. The larger the value of $\Delta$, the larger is the stability domain of the corresponding incoherent state \cite{Hakim1992}.
The most stable incoherent state, i.e., the state with the largest stability region corresponds to $\Delta=1$. In this state the population of oscillators is evenly split between two clusters having opposite phases \cite{Hakim1992}.
Fig.~\ref{fig:inc} shows the stability curves of the uniform and nonuniform incoherent states in $C_1-K$ parameter space for some values of $\tau$. For a given value of delay, the incoherent state is stable in the region below the curve and loses stability on crossing the curve into the region of higher values of $K$ and more positive values of $C_1$. 
With increase in delay, the region of stability of these states decreases since the thresholds of other collective states move towards lower values of coupling strength.
\subsection{Amplitude-Mediated Chimera State} 
\begin{figure*}
     \centering
    \includegraphics[width=0.9\textwidth]{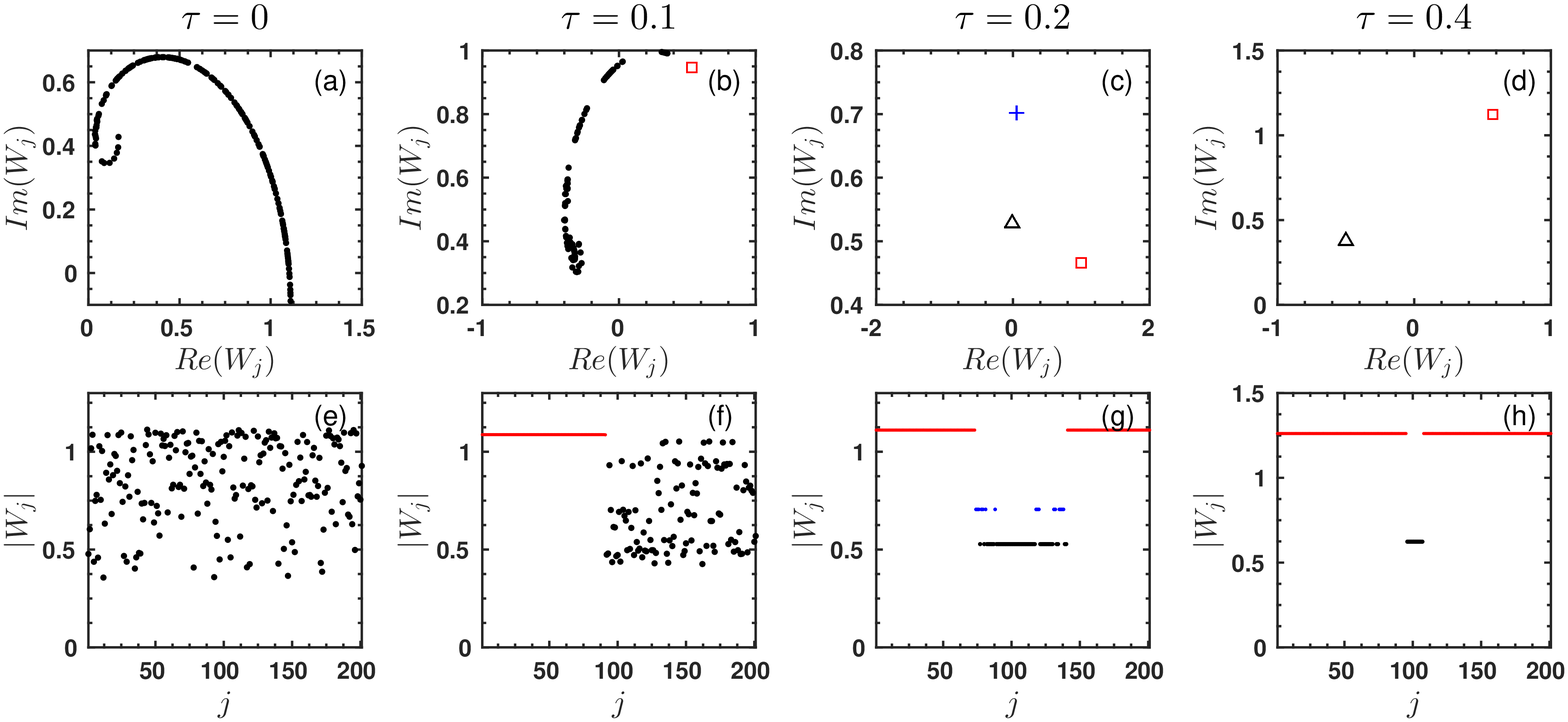}
     \caption{Snapshots of the distribution of $N=201$ oscillators in complex $W$ plane are shown in the first row of the figure for four values of time delay parameter (a) $\tau=0$, (b) $\tau=0.1$, (c) $\tau=0.2$, and (d) $\tau=0.4$. Snapshots of the profile of amplitude $|W_j|$ of the oscillators are shown in second row for (e) $\tau=0$, (f) $\tau=0.1$, (g) $\tau=0.2$, and (h) $\tau=0.4$. Other parameters are taken as $K=0.67$, $C_1=-1.75$ and $C_2=2$. With an increase in delay, the chaotic state at $\tau=0$ first goes to a chimera state ($\tau=0.1$) and then to a three-cluster state ($\tau=0.2$) and a two-cluster state ($\tau=0.4$). Each symbol (red square, black triangle and blue cross) in (c) and (d) shows the location of a point-cluster in the complex $W$ plane.}
 \label{fig:plot_amc}
  \end{figure*}

\begin{figure}[ht!]
     \centering
\includegraphics[width=0.48\textwidth]{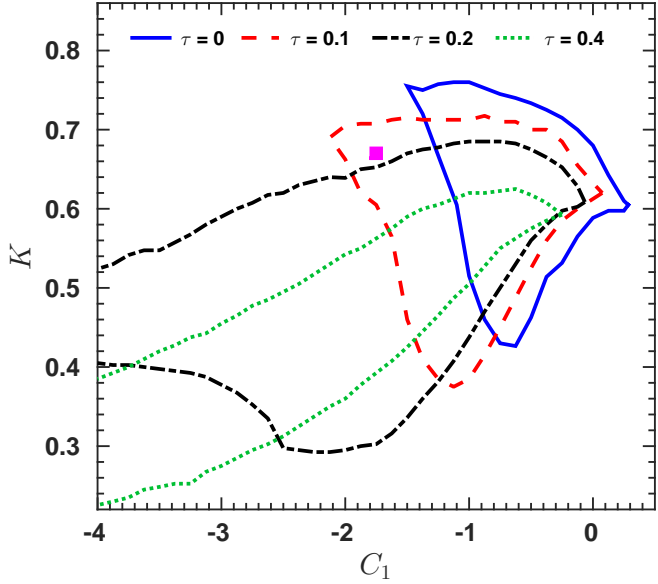}
\caption{Phase diagram for amplitude-mediated chimera (AMC) state in $C_1-K$ space for $C_2=2$ for some values of time delay parameter, $\tau=0$ (solid blue line), $\tau=0.1$ (red dashed line), $\tau=0.2$ (black dashed-dotted line), and $\tau=0.4$ (green dotted line). We have chosen a splay state as the initial condition and $N=201$. The stability region of the AMCs seems to shift towards lower values of the coupling parameter $K$ and more negative values of $C_1$. The magenta square symbol shows the data point corresponding to the parameter set of Fig.~\ref{fig:plot_amc}.}
 \label{fig:amc_delay}
  \end{figure}

Past studies on AMCs have been restricted to instantaneous interactions among the oscillators \cite{Sethia2013,Sethia2014,Rupak2018}. Here we discuss their existence and stability for finite delays in GCCGLE.
We start with a set of parameters that lies outside the existence domain of the AMC state for $\tau=0$ and numerically integrate the system of Eq.~\eqref{eq:mod}. In Fig.~\ref{fig:plot_amc}(a), we have plotted the snapshot of the distribution of 201 oscillators in complex $W$ plane for $\tau=0$ and in Fig.~\ref{fig:plot_amc}(e) we have plotted the profile of their amplitude $|W_j|$. The system is in a chaotic state characterized by the somewhat $\rho$-shaped distribution (Fig.~\ref{fig:plot_amc}(a)) which rotates and deforms as a function of time. 
As we increase delay, the system goes to the AMC state as shown in Fig.~\ref{fig:plot_amc}(b) and Fig.~\ref{fig:plot_amc}(f). Fig.~\ref{fig:plot_amc}(b) shows the distribution in the complex plane and Fig.~\ref{fig:plot_amc}(f) shows the amplitude-profile of the AMC state for $\tau=0.1$. In the AMC state, a fraction of the oscillators behave coherently (shown by a red square in Fig.~\ref{fig:plot_amc}(b)) while the rest of the oscillators show incoherent behavior (shown by black dots in Fig.~\ref{fig:plot_amc}(b)). As time delay is further increased, the system first goes to a three-cluster state (shown for $\tau=0.2$ in Fig.~\ref{fig:plot_amc}(c) and Fig.~\ref{fig:plot_amc}(g)) and then to a two-cluster state (shown for $\tau=0.4$ in Fig.~\ref{fig:plot_amc}(d) and Fig.~\ref{fig:plot_amc}(h)).
The results plotted together in Fig.~\ref{fig:plot_amc} show that the presence of time delay can modify the existence domain of the AMC state. A set of parameters can lie outside or inside the existence domain of the AMC state depending on the value of the delay parameter.
In order to better understand how an increase in delay affects the existence domain of AMCs, we have plotted their existence region in $C_1-K$ space for a fixed value of $C_2=2$ in Fig.~\ref{fig:amc_delay} for various values of time delay parameter $\tau$. We have taken the initial condition to be a splay state. 
The numerical results indicate that the region of stability of AMCs shifts towards lower values of coupling strength $K$ and more negative values of $C_1$. We will show later for small delays that this shift is similar to what one would get on adding a nonlinear global coupling term to the non-delayed GCCGLE.
\section{Small delay limit}\label{sec:small_delay}
\begin{figure*}
     \centering
\includegraphics[width=0.9\textwidth]{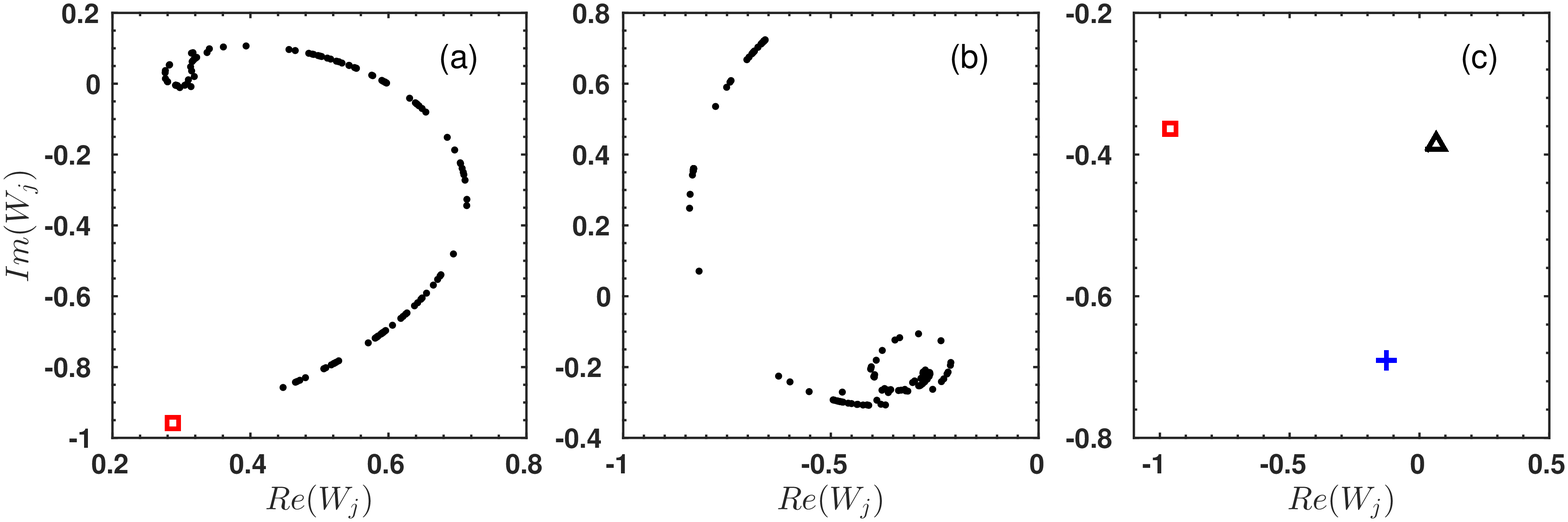} 
     \caption{(a) The system is in a chimera state for $\tau=0$. Red square represents the coherent oscillators and black dots represent the incoherent oscillators. (b) For $\tau=0.1$, the system of Eq.~\eqref{eq:expansion} goes towards a chaotic state if we ignore the nonlinear global coupling term.
     (c) The system goes to three-cluster state on including the nonlinear global coupling term when $\tau=0.1$. Each symbol (red square, black triangle and blue cross) shows the location of a point-cluster in the complex $W$ plane. The initial condition is a splay state and parameters are $C_2=2$, $C_1=-1.3$, $K=0.713$, and $N=201$.}  
  \label{fig:exp_plt}
  \end{figure*}

In order to understand the effects emerging from the presence of time delay, we Taylor expand Eq.~\eqref{eq:mod} for small delays.
For small lags, we can take $W_j(t-\tau)=W_j(t)-\tau \dot{W_j}(t)$ and $\overline{W}(t-\tau)=\overline{W}(t)-\tau \dot{\overline{W}}(t)$, where we ignore $O(\tau^2)$ and higher order terms.
On substituting these in Eq.~\eqref{eq:mod} and after re-arranging some terms, we can write the set of model equations as
\begin{eqnarray}\label{eq:expansion}
\dot{W}_j(t)=(1-K(1+i C_1))W_j(t)-(1+i C_2)|W_j(t)|^2 W_j(t)\nonumber\\
+\underbrace{K(1-\tau)(1+i C_1)\overline{W}(t)}_{I}\;\;\;\;\;\;\;\;\;\;\;\;\;\;\;\;\;\;\;\;\;\nonumber\\
+\underbrace{\tau K(1+i C_1)(1+i C_2)\frac{1}{N} \sum_{m=1}^N |W_m|^2 W_m}_{II},\;\;\;\;\;\;\;
\end{eqnarray}
j=1,..,N. We see that one outcome of the presence of time delay is the reduction of the contribution of the linear global coupling term $I$ from $K$ to $(1-\tau)K$.
Another outcome of the presence of time delay is the addition of a nonlinear global coupling term $II$ to the equation. To see how each of these effects modifies the collective behavior of the system, we
numerically integrate Eq.~\eqref{eq:expansion} for a set of parameters for which the system is in a chimera state (Fig.~\ref{fig:exp_plt}(a)) in the absence of delay (i.e., $\tau=0$). For a finite delay, if we ignore the nonlinear global coupling term $II$, we find that the system goes from chimera state (Fig.~\ref{fig:exp_plt}(a)) towards chaos (Fig.~\ref{fig:exp_plt}(b)) as we increase $\tau$.  This is a result of the reduction of linear global coupling by a factor of $(1-\tau)$ in term $I$. However, if we retain the nonlinear global coupling term, the system goes from chimera (Fig.~\ref{fig:exp_plt}(a)) to three-cluster state (Fig.~\ref{fig:exp_plt}(c)). 
This implies that the nonlinear global coupling term increases the effective coupling strength of the system and this effect is more dominating than the reduction in the linear global coupling. 
Therefore, it seems that the main effect of introducing time delay in interaction is to introduce a nonlinear global coupling term in the system which drives the system with a higher mean-field strength (i.e., which increases the effective coupling strength of the system).
Hence the thresholds of various collective states move towards lower values of coupling parameter $K$ with an increase in delay.

\begin{figure}[ht]
     \centering
\includegraphics[width=0.49\textwidth]{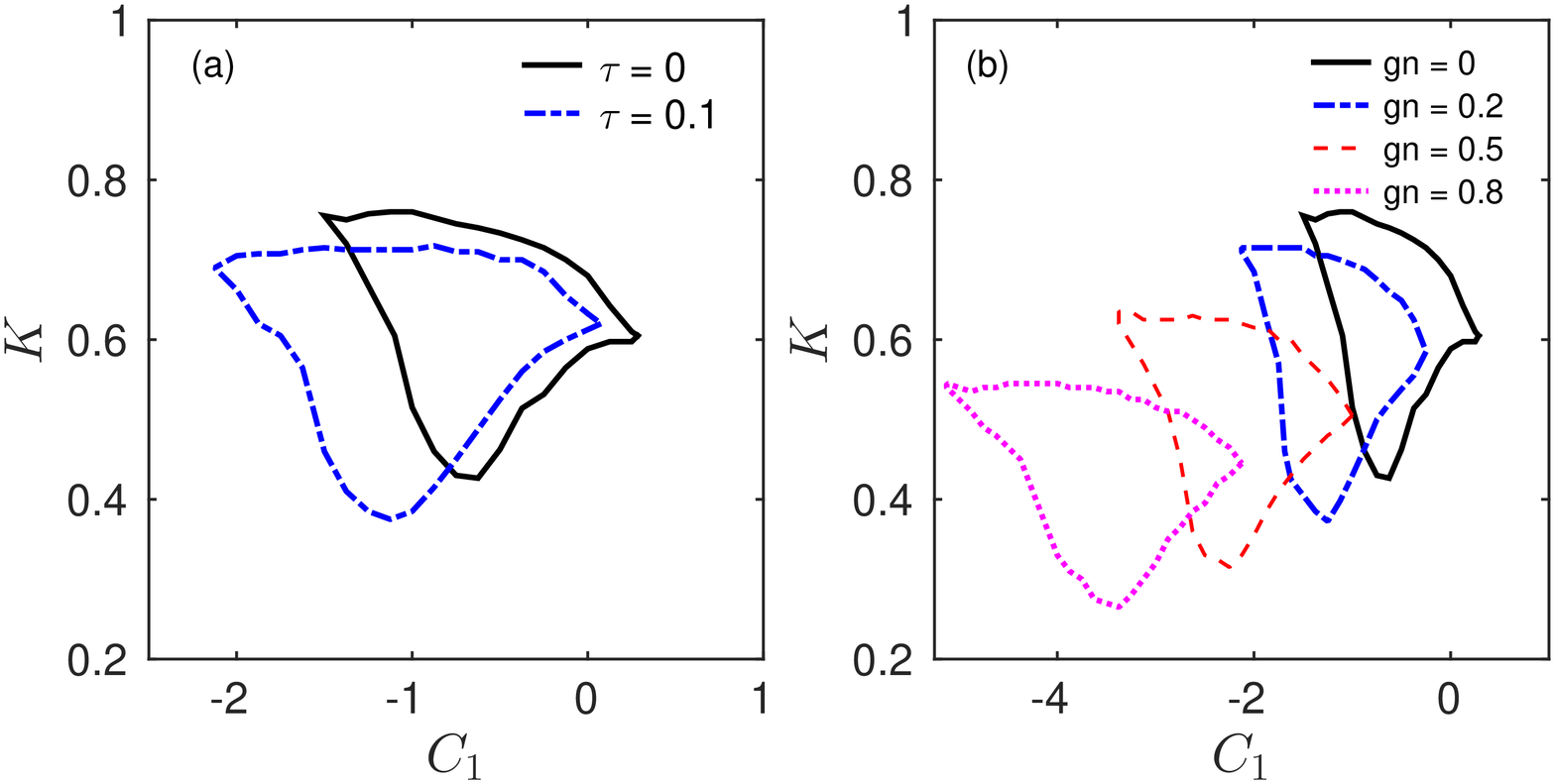}
     \caption{The region of existence of amplitude-mediated chimeras for various values of (a) $\tau$, the delay parameter and (b) $gn$, the strength of the nonlinear global coupling term . We have taken $C_2=2$, $N=100$ and a splay state as the initial condition for the simulations.}
  \label{fig:gn_plt}
  \end{figure}

To further support this argument we consider the non-delayed system that has a nonlinear global coupling term in addition to the linear coupling, such that 
\begin{eqnarray}\label{eq:nonlinear_coup}
\dot{W}_j(t)=(1-K(1+i C_1))W_j(t)-(1+i C_2)|W_j(t)|^2 W_j(t)\nonumber\\
+K(1+i C_1)\overline{W}(t) + gn\;
K (1+i C_2)\frac{1}{N} \sum_{m} |W_m|^2 W_m,\;\;\;\;\;\;
\end{eqnarray}
where $gn$ is the strength of the nonlinear global coupling term. In Fig.~\ref{fig:gn_plt} (b), we have shown the region of existence of AMCs for various values of $gn$ for a given set of initial conditions. We find that as we increase the contribution of the nonlinear global coupling term, the region of stability moves towards lower $K$ and more negative $C_1$, similar to what we observe in the time-delayed system with an increase in delay (Fig.~\ref{fig:gn_plt} (a) (obtained from numerical integration of Eq.~\eqref{eq:mod})).
The effect of nonlinear global coupling on the system of globally coupled Stuart-Landau oscillators has been studied recently by Schmidt and Krischer \cite{Schmidt2015a}.
Similar nonlinear global coupling was also considered by the same authors in a system of complex Ginzburg-Landau equation \cite{Schmidt2015}.
\section{Discussion}\label{sec:discussion}
In this paper, we have studied the changes in the regions of existence and stability of different collective states of a system of globally coupled complex Ginzburg Landau oscillators on introduction of time delay in interactions.
The various collective states observed are the single-cluster (homogeneous limit-cycle) state, multicluster states (two point- and three point-clusters), amplitude-mediated chimera state, chaotic state and the incoherent state. These states are discussed in the order of their appearance with decreasing coupling strength $K$.
We have derived the analytical stability curves wherever possible and have given the numerical stability curves when we do not have an exact analytical expression.
We find that an increase in time delay seems to favor the stabilization of the single-cluster state by decreasing the threshold of its stability towards lower values of the coupling parameter. Time delay also seems to suppress the oscillatory instability in two-cluster quasi-periodic state.
With increase in delay, the stability region of amplitude-mediated chimeras shifts towards lower values of $K$ and more negative values of $C_1$ and seems to continue into the region with quite high negative values of $C_1$. 
The region of stability of the incoherent state is seen to decrease monotonically with increase in delay.
The detailed effect of time delay on the chaotic state remains to be investigated.
The overall effect of time-delay is found to be the shifting of the region of stability of various collective states towards lower values of $K$ and more negative values of $C_1$. We find that for small delays this effect is similar to adding a nonlinear global coupling term to the non-delayed system.


\begin{thebibliography}{28}
\expandafter\ifx\csname natexlab\endcsname\relax\def\natexlab#1{#1}\fi
\expandafter\ifx\csname bibnamefont\endcsname\relax
  \def\bibnamefont#1{#1}\fi
\expandafter\ifx\csname bibfnamefont\endcsname\relax
  \def\bibfnamefont#1{#1}\fi
\expandafter\ifx\csname citenamefont\endcsname\relax
  \def\citenamefont#1{#1}\fi
\expandafter\ifx\csname url\endcsname\relax
  \def\url#1{\texttt{#1}}\fi
\expandafter\ifx\csname urlprefix\endcsname\relax\def\urlprefix{URL }\fi
\providecommand{\bibinfo}[2]{#2}
\providecommand{\eprint}[2][]{\url{#2}}


\bibitem{Newell1969}
  \bibinfo{author}{\bibfnamefont{A.~C.} \bibnamefont{Newell}},
   \bibnamefont{and}  \bibinfo{author}{\bibfnamefont{J.~A.} \bibnamefont{Whitehead}},
   \emph{\bibinfo{title}{Finite bandwidth, finite amplitude convection}},
   \bibinfo{journal}{Journal of Fluid Mechanics}
  \textbf{\bibinfo{volume}{38}}, \bibinfo{pages}{279--303} (\bibinfo{year}{1969}).


\bibitem{Hakim1992}
  \bibinfo{author}{\bibfnamefont{V.} \bibnamefont{Hakim}},
  \bibnamefont{and} \bibinfo{author}{\bibfnamefont{W.-~J.} \bibnamefont{Rappel}},
  \emph{\bibinfo{title}{Dynamics of the globally coupled complex Ginzburg-Landau equation}},
   \bibinfo{journal}{Phys. Rev. A}
  \textbf{\bibinfo{volume}{46}}, \bibinfo{pages}{R7347} (\bibinfo{year}{1992}).
  
\bibitem{Willaime1991a}
  \bibinfo{author}{\bibfnamefont{H.} \bibnamefont{Willaime}},
    \bibinfo{author}{\bibfnamefont{O.} \bibnamefont{Cardoso}},
  \bibnamefont{and} \bibinfo{author}{\bibfnamefont{P.} \bibnamefont{Tabeling}},
  \emph{\bibinfo{title}{Coupled oscillators-An accurate model for describing the dynamics of lines of vortices}},
   \bibinfo{journal}{Eur. J. Mech. B Fluids}
  \textbf{\bibinfo{volume}{10}}, \bibinfo{pages}{165--171} (\bibinfo{year}{1991}).

\bibitem{Willaime1991b}
  \bibinfo{author}{\bibfnamefont{H.} \bibnamefont{Willaime}},
    \bibinfo{author}{\bibfnamefont{O.} \bibnamefont{Cardoso}},
  \bibnamefont{and} \bibinfo{author}{\bibfnamefont{P.} \bibnamefont{Tabeling}},
  \emph{\bibinfo{title}{Frustration in a linear array of vortices}},
   \bibinfo{journal}{Phys. Rev. Lett.}
  \textbf{\bibinfo{volume}{67}}, \bibinfo{pages}{3247} (\bibinfo{year}{1991}).
  
  
\bibitem{Nakagawa1993}
  \bibinfo{author}{\bibfnamefont{N.} \bibnamefont{Nakagawa}},
  \bibnamefont{and} \bibinfo{author}{\bibfnamefont{Y.} \bibnamefont{Kuramoto}},
  \emph{\bibinfo{title}{Collective chaos in a population of globally coupled oscillators}},
   \bibinfo{journal}{Prog. Theor. Phys.}
  \textbf{\bibinfo{volume}{89}}, \bibinfo{pages}{313--323} (\bibinfo{year}{1993}).

\bibitem{Nakagawa1994}
  \bibinfo{author}{\bibfnamefont{N.} \bibnamefont{Nakagawa}},
  \bibnamefont{and} \bibinfo{author}{\bibfnamefont{Y.} \bibnamefont{Kuramoto}},
  \emph{\bibinfo{title}{From collective oscillations to collective chaos in a globally coupled oscillator system}},
   \bibinfo{journal}{Physica D}
  \textbf{\bibinfo{volume}{75}}, \bibinfo{pages}{74--80} (\bibinfo{year}{1994}).

\bibitem{Nakagawa1995}
  \bibinfo{author}{\bibfnamefont{N.} \bibnamefont{Nakagawa}},
  \bibnamefont{and} \bibinfo{author}{\bibfnamefont{Y.} \bibnamefont{Kuramoto}},
  \emph{\bibinfo{title}{Anomalous Lyapunov spectrum in globally coupled oscillators}},
   \bibinfo{journal}{Physica D}
  \textbf{\bibinfo{volume}{80}}, \bibinfo{pages}{307--316} (\bibinfo{year}{1995}).


\bibitem{Hakim1994}
  \bibinfo{author}{\bibfnamefont{V.} \bibnamefont{Hakim}},
  \bibnamefont{and} \bibinfo{author}{\bibfnamefont{W.-~J.} \bibnamefont{Rappel}},
  \emph{\bibinfo{title}{Noise-induced periodic behaviour in the globally coupled complex Ginzburg-Landau equation}},
   \bibinfo{journal}{Europhys. Lett.}
  \textbf{\bibinfo{volume}{27}}, \bibinfo{pages}{637--642} (\bibinfo{year}{1994}).
  
\bibitem{Chabanol1997}
\bibinfo{author}{\bibfnamefont{M.-~L.} \bibnamefont{Chabanol}},
  \bibinfo{author}{\bibfnamefont{V.} \bibnamefont{Hakim}},
  \bibnamefont{and} \bibinfo{author}{\bibfnamefont{W.-~J.} \bibnamefont{Rappel}},
  \emph{\bibinfo{title}{Collective chaos and noise in the globally coupled complex Ginzburg-Landau equation}},
   \bibinfo{journal}{Physica D}
  \textbf{\bibinfo{volume}{103}}, \bibinfo{pages}{273--293} (\bibinfo{year}{1997}).  


\bibitem{Sethia2013}
\bibinfo{author}{\bibfnamefont{G.~C.} \bibnamefont{Sethia}},
\bibinfo{author}{\bibfnamefont{A.} \bibnamefont{Sen}},
 \bibnamefont{and} \bibinfo{author}{\bibfnamefont{G.~L.} \bibnamefont{Johnston}},
  \emph{\bibinfo{title}{Amplitude-mediated chimera states}},
   \bibinfo{journal}{Phys. Rev. E}
  \textbf{\bibinfo{volume}{88}}, \bibinfo{pages}{042917} (\bibinfo{year}{2013}).  
  
\bibitem{Sethia2014}
\bibinfo{author}{\bibfnamefont{G.~C.} \bibnamefont{Sethia}},
 \bibnamefont{and} \bibinfo{author}{\bibfnamefont{A.} \bibnamefont{Sen}},
  \emph{\bibinfo{title}{Chimera States: The Existence Criteria Revisited}},
   \bibinfo{journal}{Phys. Rev. Lett.}
  \textbf{\bibinfo{volume}{112}}, \bibinfo{pages}{144101} (\bibinfo{year}{2014}).

\bibitem{Zakharova2016}
\bibinfo{author}{\bibfnamefont{A.} \bibnamefont{Zakharova}},
\bibinfo{author}{\bibfnamefont{M.} \bibnamefont{Kapeller}},
 \bibnamefont{and} \bibinfo{author}{\bibfnamefont{E.} \bibnamefont{Sch\"{o}ll}},
  \emph{\bibinfo{title}{Amplitude chimeras and chimera death in dynamical networks}},
   \bibinfo{journal}{Journal of Physics: Conference Series}
  \textbf{\bibinfo{volume}{727}}, \bibinfo{pages}{012018} (\bibinfo{year}{2016}).  

%

\bibitem{Falcke1995}
\bibinfo{author}{\bibfnamefont{M.} \bibnamefont{Falcke}},
\bibinfo{author}{\bibfnamefont{H.} \bibnamefont{Engel}},
 \bibnamefont{and} \bibinfo{author}{\bibfnamefont{M.} \bibnamefont{Neufeld}},
  \emph{\bibinfo{title}{Cluster formation, standing waves, and stripe patterns in oscillatory active media with local and global coupling}},
   \bibinfo{journal}{Phys. Rev. E}
  \textbf{\bibinfo{volume}{52}}, \bibinfo{pages}{763--771} (\bibinfo{year}{1995}).
  
  
\bibitem{Nakao1999}
\bibinfo{author}{\bibfnamefont{H.} \bibnamefont{Nakao}},
  \emph{\bibinfo{title}{Anomalous spatio-temporal chaos in a two-dimensional system of nonlocally coupled oscillators}},
   \bibinfo{journal}{Chaos}
  \textbf{\bibinfo{volume}{9}}, \bibinfo{pages}{902--909} (\bibinfo{year}{1999}). 
  
 
\bibitem{Battogtokh2000}
\bibinfo{author}{\bibfnamefont{D.} \bibnamefont{Battogtokh}},
 \bibnamefont{and} \bibinfo{author}{\bibfnamefont{Y.} \bibnamefont{Kuramoto}},
  \emph{\bibinfo{title}{Turbulence of nonlocally coupled oscillators in the Benjamin-Feir stable regime}},
   \bibinfo{journal}{Phys. Rev. E}
  \textbf{\bibinfo{volume}{61}}, \bibinfo{pages}{3227--3229} (\bibinfo{year}{2000}). 
  
  
\bibitem{Kuramoto2002}
\bibinfo{author}{\bibfnamefont{Y.} \bibnamefont{Kuramoto}},
 \bibnamefont{and} \bibinfo{author}{\bibfnamefont{D.} \bibnamefont{Battogtokh}},
  \emph{\bibinfo{title}{Coexistence of coherence and incoherence in nonlocally coupled phase oscillators}},
   \bibinfo{journal}{Nonlin. Phenom. Compl. Sys.}
  \textbf{\bibinfo{volume}{5}}, \bibinfo{pages}{380} (\bibinfo{year}{2002}).


\bibitem{Kemeth2018}
\bibinfo{author}{\bibfnamefont{F.~P.} \bibnamefont{Kemeth}},
\bibinfo{author}{\bibfnamefont{S.~W.} \bibnamefont{Haugland}},
 \bibnamefont{and} \bibinfo{author}{\bibfnamefont{K.} \bibnamefont{Krischer}},
  \emph{\bibinfo{title}{Cluster singularities: the unfolding of clustering behavior in globally coupled oscillatory systems}},
   \bibinfo{journal}{arXiv}
  \textbf{\bibinfo{volume}{1807}}, \bibinfo{pages}{11231} (\bibinfo{year}{2018}).

   
\bibitem{Rupak2018}
\bibinfo{author}{\bibfnamefont{R.} \bibnamefont{Mukherjee}},
 \bibnamefont{and} \bibinfo{author}{\bibfnamefont{A.} \bibnamefont{Sen}},
  \emph{\bibinfo{title}{Amplitude mediated chimera states with active and inactive oscillators}},
   \bibinfo{journal}{Chaos}
  \textbf{\bibinfo{volume}{28}}, \bibinfo{pages}{053109} (\bibinfo{year}{2018}).  

\bibitem{Schmidt2015a}
\bibinfo{author}{\bibfnamefont{L.} \bibnamefont{Schmidt}},
 \bibnamefont{and} \bibinfo{author}{\bibfnamefont{K.} \bibnamefont{Krischer}},
  \emph{\bibinfo{title}{Clustering as a prerequisite for chimera states in globally coupled systems}},
   \bibinfo{journal}{Phys. Rev. Lett.}
  \textbf{\bibinfo{volume}{114}}, \bibinfo{pages}{034101} (\bibinfo{year}{2015}).    
  
  
\bibitem{Schmidt2015}
\bibinfo{author}{\bibfnamefont{L.} \bibnamefont{Schmidt}},
 \bibnamefont{and} \bibinfo{author}{\bibfnamefont{K.} \bibnamefont{Krischer}},
  \emph{\bibinfo{title}{Chimeras in globally coupled oscillatory systems: From ensembles of oscillators to spatially continuous media}},
   \bibinfo{journal}{Chaos}
  \textbf{\bibinfo{volume}{25}}, \bibinfo{pages}{064401} (\bibinfo{year}{2015}).  

  
  
\end{thebibliography}
\end{document}